\documentclass[aps,twocolumn]{revtex4}
\pdfoutput=1

\usepackage{amsmath,amssymb,graphicx}
\usepackage{algorithmic}
\usepackage{enumerate}
\usepackage{color}

\begin{document}
\title{Simulating thermal boundary conditions of spin-lattice models with weighted averages}
\author{Wenlong Wang}
\email{wenlong@physics.umass.edu}
\affiliation{Department of Physics, University of Massachusetts,
Amherst, Massachusetts 01003 USA}
\affiliation{Department of Physics and Astronomy, Texas A$\&$M University,
College Station, Texas 77843-4242, USA}

\begin{abstract}
Thermal boundary conditions has played an increasingly important role in revealing the nature of short-range spin glasses and is likely to be relevant also for other disordered systems. Diffusion method initializing each replica with a random boundary condition at the infinite temperature using population annealing has been used in recent large-scale simulations. However, the efficiency of this method can be greatly suppressed because of temperature chaos. For example, most samples have some boundary conditions that are completely eliminated from the population in the process of annealing at low temperatures. In this work, I study a weighted average method to solve this problem by simulating each boundary conditions separately and collect data using weighted averages. The efficiency of the two methods are studied using both population annealing and parallel tempering, showing that the weighted average method is more efficient and accurate.
\end{abstract}

\maketitle

\section{INTRODUCTION}
Thermal boundary conditions (TBC) includes the set of all combinations of periodic/antiperiodic boundary conditions in each spatial dimension \cite{TBC,TC,BC}. For example in three dimensions ($d = 3$), there are $2^d=8$ possible choices. Each boundary condition $i$ occurs in the thermal ensemble with weight $p_i$ depends on its free energy $F_i$ as $p_i = \exp[-\beta(F_i-F)]$, where $F$ is the total free energy of the system in TBC and $\beta$ is the inverse temperature. TBC was initially motivated to reduce domain-wall effects in spin glasses \cite{TBC} and was later shown to be useful in studying temperature chaos and bond chaos via boundary condition crossings, i.e, the weights $\{p_i\}$ change chaotically as a function of temperature or couplings \cite{TC,BC}. Previously, thermal boundary conditions was used with exact algorithms for finding ground states of two-dimensional spin glasses \cite{landry:02,thomas:07} (referred to as ``extended'' boundary conditions). A restricted version of thermal boundary conditions using periodic and anti-periodic boundary conditions in a single direction (a subset of TBC) was also used in Refs.~\cite{hukushima:99,sasaki:05,sasaki:07b,hasenbusch:93}.

Simulating the full TBC ensemble in Refs.~\cite{TBC,TC,BC} was done using population annealing (PA)~\cite{hukushima:03,machta:10,machta:11,zhou:10}. See Ref.~\cite{pamc} for a recent discussion of the algorithm. Population annealing works with a population of replicas and aims to maintain thermal equilibrium while lowering the temperature. When temperature is decreased, the population is resampled according to the Boltzmann weights of the replicas, followed by regular Monte Carlo sweeps to all replicas in the population. In this work, the Metropolis algorithm is used. Simulating TBC using the diffusion method with PA is very simple. One can initialize each replica with a random boundary condition at $\beta=0$ and then boundary conditions are resampled along with the replicas. The word ``diffusion'' would become apparent when one implements the method in parallel tempering, as discussed in Sec.~\ref{diffusion}.

It was noticed in Refs.~\cite{TBC,TC} that the efficiency of the diffusion method can be greatly suppressed by temperature chaos. Some boundary conditions can be totally removed from the population. This is not satisfactory as there is no mechanism to recover these lost boundary conditions once they are eliminated from the population. Furthermore, temperature chaos predicts that these boundary conditions could become important again at lower temperatures. Therefore, it is worth to study a new method that does not have this problem. The most natural way is to simulate each boundary condition separately and combine the observables using weighted averages with free energy. How to weight different kinds of observables is discussed in Sec.~\ref{wa}. Since it is also interesting to study the performance of parallel tempering (PT) in simulating the full set of thermal boundary conditions, the two methods are therefore studied in both PA and PT.

The paper is organized as follows: Section~\ref{model} introduces the Edwards-Anderson model, measured quantities, the diffusion method and the weighted average method. Numerical results are shown in Sec.~\ref{result}, followed by concluding remarks and future challenges in Sec.~\ref{conclusion}.

\section{MODELS, OBSERVABLES AND METHODS}
\label{model}
\subsection{The Edwards-Anderson model}
The Edwards-Anderson (EA) Hamiltonian is defined as
\begin{equation}
H=-\sum\limits_{\langle ij \rangle} J_{ij} S_i S_j,
\end{equation}
where $\{S_i=\pm1\}$ are the spin degrees of freedom defined on a three-dimensional cubic lattice. The sum over $\langle ij \rangle$ means sum over all nearest neighbours. $J_{ij}$ is the coupling between spins $S_i$ and $S_j$ and all couplings are independently chosen from the standard Gaussian distribution with mean 0 and variance 1. Thermal boundary conditions is applied to all instances.

\subsection{Observables}
There are three classes of observables that need to be treated differently using weighted averages. The first class are observables that are functions of a single replica like the energy $E$ or the magnetization $m$. The second class are the thermodynamic observables of the free energy $F$ and the entropy $S$. There is a third class of observables in spin glasses due to the nature of the symmetry breaking which are functions of two replicas like the spin overlap $q$ defined as
\begin{equation}
q_{a b}=\dfrac{1}{N}\sum\limits_i S_i^a S_i^b,
\end{equation}
where micro-states $a,b$ are chosen independently from the Boltzmann distribution. Note that $a,b$ can be chosen from the same boundary conditions or different boundary conditions. Another quantity in this class is the link overlap which needs some care in the definition due to the change of boundary conditions, the link overlap $q_\ell$ is defined as
\begin{equation}
{q_\ell}_{a b}=\dfrac{1}{dN}\sum\limits_{\langle ij \rangle} S^a_i \mathrm{sign}(J^a_{ij}) S^a_j S^b_i \mathrm{sign}(J^b_{ij}) S^b_j,
\end{equation}
where the sign function is $\pm 1$ depends on whether the argument is positive or negative respectively. The definition has nothing to do with the weighted average method, but this is a more useful definition in the TBC setting for studying the nature of spin glasses. In this way, the difference of the link overlap of two different boundary conditions arises only from the domain walls, not from a mixture of domain walls and boundary condition differences. In the following, I will focus on the spin overlap function and will refer to it as the overlap function where no confusion arises. Note also that there is a $8\times 8$ overlap function matrix in the TBC ensemble. An important statistic $I$, which quantifies the weight of an overlap distribution near $q=0$, is defined as
\begin{equation}
I=\int_{-0.2}^{0.2} P(q) dq,
\end{equation}
where $P(q)$ is an overlap distribution function.

\subsection{The diffusion method}
\label{diffusion}
We have already discussed the diffusion method in population annealing, which initializes each replica with a random boundary condition at $\beta=0$. For completeness, I also introduce and study the diffusion method in parallel tempering because parallel tempering is widely used in spin-glass simulations. Following the same strategy, one can simulate the TBC using parallel tempering by generating random states with random boundary conditions at $\beta=0$. Then the boundary conditions diffuse along with the replicas under the swap moves of parallel tempering, hence the name the diffusion method. The convenience of working with $\beta=0$ is because proposing a boundary condition change at finite temperatures can be costly as many bonds are affected when boundary condition changes. Furthermore, this is also essential to measure the absolute free energy.

The implementation of this method is simple and detailed balance is satisfied. However, the efficiency of the method still needs to be studied. Since in this method, PA and PT work in a similar way, one may expect that PT should suffer from temperature chaos too. This turns out to be the case as discussed in Secs.~\ref{m1} and \ref{limitation}. The effect of boundary condition eliminations in PA is replaced by one or more diffusion bottlenecks in PT. In the next, I discuss the weighted average method, which does not have this problem.

\subsection{The weighted average method}
\label{wa}
It was shown that the absolute free energy can be measured very accurately using the free energy perturbation method in both population annealing and parallel tempering ~\cite{machta:10,PTFE}. It can be shown from statistical mechanics that the average energy, entropy, free energy and the spin overlap distribution should be averaged as
\begin{eqnarray}
E &=& \sum_i E_i p_i, \\
S &=& \sum_i S_i p_i -\sum_i p_i \log p_i, \\
F &=& \sum_i F_i p_i + T \sum_i p_i \log p_i, \\
P(q)&=&\sum_{ij} P_{ij}(q) p_i p_j,
\end{eqnarray}
where $p_i=\dfrac{e^{-\beta F_i}}{\sum\limits_{j} e^{-\beta F_j}}$, $E_i, S_i$ and $F_i$ are the energy, entropy and free energy of boundary condition $i$ and $P_{ij}(q)$ is the overlap distribution between the boundary conditions $i$ and $j$.

It is worth noting that the weighted average method generates a lot more information than the diffusion method. For example, the overlap matrix is briefly discussed in Sec.~\ref{conclusion}. For now, we discuss briefly of the implementation of the weighted average method in PA and PT. In the weighted average method, each boundary condition is simulated independently. It is only when computing the spin overlaps that one needs communications between replicas with different boundary conditions. Therefore, in this way, PT can be simulated using parallel computing too. Since it is usually a practice to simulated two independent sets of replicas of each boundary condition, one can use 8 or 16 threads in the simulation. This is not doable in the diffusion method of PT. On the other hand, PA is intrinsically parallel and is also much more flexible with the number of threads. In my simulations, I have used OpenMP parallel computing for both PA and PT. For the equilibration measure of PA, one can either use the weighted average or the minimum of the entropy of families \cite{pamc} of all boundary conditions.

%\begin{table}
%\caption{
%Summary of strengths (normal) and weaknesses (italic) of the diffusion method (M1) and the weighted average method (M2) in PA and PT. The weaknesses are the strengths that are missing. One can use a combination of simulation methods in a large-scale TBC simulations.
%\label{table}
%}
%\begin{tabular*}{\columnwidth}{@{\extracolsep{\fill}} l c l}
%\hline
%\hline
%Methods  & PA & PT\\
%\hline
%M1  &simple to implement &simple to implement\\
%&simple instances &simple instances\\
%&massively parallel &parallel to some extent\\
%&\textit{large memory} &small memory\\
%& &\textit{less parallel}\\
%M2  &hard chaotic instances &hard chaotic instances \\
%&massively parallel &small memory\\
%&\textit{large memory} &\textit{limited parallel}\\
%\hline
%\hline
%\end{tabular*}
%\end{table}

\section{Results}
\label{result}
In this section, I investigate the performance of the diffusion method and the weighted average method. I first compare PA and PT for the diffusion method in Sec.~\ref{m1}. The conclusion is that the diffusion method, while works but can be inefficient and the limitations of the method are discussed in Sec.~\ref{limitation}. Finally, I study the performance of the weighted average method, compare PA and PT and also with the diffusion method, showing that the weighted average method works better than the diffusion method in Sec.~\ref{m2}.

\subsection{PA and PT: the diffusion method}
\label{m1}
In this section, I compare the efficiency of the diffusion method of PA and PT in simulating thermal boundary conditions. The large-scale population annealing data is taken from a recent work on temperature chaos \cite{TC}. In the implementation of parallel tempering, I used $N_{\beta}$ temperatures evenly distributed in $\beta$ for the high temperature part and $N_T$ temperatures evenly distributed in $T$ for the low temperature part. No Monte Carlo sweeps were done at $\beta=0$, it is sufficient to generate random states and boundary conditions at the infinite temperature. The amount of work in this work is counted in Monte Carlo sweeps and one Monte Carlo sweep is a sequential update of all the spins for a replica at a temperature once. The work is distributed evenly between thermalization and data collection. The simulation parameters of PA and PT are summarized in Tables~\ref{para1} and \ref{para2} respectively. The main conclusion of this section is that PA and PT has similar efficiency in the diffusion method. The limitations of the method are discussed in Sec.~\ref{limitation}.

\begin{table}
\caption{
The simulation parameters of PA \cite{TC} using the diffusion method (DF) for different system sizes $L$ with thermal boundary conditions. $R$ represents the total number of replicas, $1/\beta_{\rm{max}}$ is the lowest temperature
simulated, $N_T$ the number of temperatures in the annealing
schedule evenly distributed in $\beta$, $N_S$ the number of sweeps per replica per temperature and $M$ the number
of samples. The simulation parameters using the weighted average method (WA) are also shown in the table, where $R$ represents the population size of each boundary condition.
\label{para1}
}
\begin{tabular*}{\columnwidth}{@{\extracolsep{\fill}} l c c c c c c}
\hline
\hline
Method &$L$  &$R$ & $1/\beta_{\rm{max}}$ & $N_T$ & $N_S$ & $M$ \\
\hline
DF &$4$  &$5\, 10^4$ & $0.2$     & $101$ & $10$  & $101$ \\
DF &$6$  &$2\, 10^5$ & $0.2$     & $101$ & $10$  & $101$ \\
DF &$8$  &$5\, 10^5$ & $0.2$     & $201$ & $10$  & $101$ \\
DF &$10$ &$2\, 10^6$ & $0.2$     & $301$ & $10$  & $101$ \\
WA &$4$  &$5\, 10^3$ & $0.2$     & $101$ & $10$  & $101$ \\
WA &$6$  &$2\, 10^4$ & $0.2$     & $101$ & $10$  & $101$ \\
WA &$8$  &$5\, 10^4$ & $0.2$     & $201$ & $10$  & $101$ \\
WA &$10$ &$2\, 10^5$ & $0.2$     & $301$ & $10$  & $101$ \\
WA &$12$ &$2\, 10^5$ & $1/3$     & $301$ & $10$  & $101$ \\
\hline
\hline
\end{tabular*}
\end{table}

\begin{table}
\caption{
The simulation parameters of PT using the diffusion method (DF) for different system sizes $L$ with thermal boundary conditions. $N_{\beta}$ represents the number of temperatures in the high temperature part uniformly distributed in $\beta$, $N_T$ the number of temperatures in the low temperature part uniformly distributed in $T$, $T_{\rm{min}}$ the lowest temperature simulated, $T_{\rm{max}}$ the highest temperature simulated in the low temperature part, $N_S$ the number of sweeps per temperature and $M$ the number of samples. The work is evenly distributed in two independent sets of replicas. The simulation parameters using the weighted average method (WA) are also shown in the table, where $N_S$ represents the number of sweeps of each boundary condition.
\label{para2}
}
\begin{tabular*}{\columnwidth}{@{\extracolsep{\fill}} l c c c c c c c}
\hline
\hline
Method &$L$  &$N_{\beta}$ & $N_T$ &$T_{\rm{min}}$ &$T_{\rm{max}}$ & $N_S$ & $M$ \\
\hline
DF &$4$  & $5$ &10 & $0.2$     & $2.0$ & $2\,10^6$  & $101$ \\
DF &$6$  & $5$ &20 & $0.2$     & $2.0$ & $4\,10^6$  & $101$ \\
DF &$8$  & $10$ &30 & $0.2$     & $2.0$ & $1.2\,10^7$  & $101$ \\
DF &$10$ & $10$ &40 & $0.2$     & $2.0$ & $6\,10^7$  & $101$ \\
WA &$4$  & $5$ &10 & $0.2$     & $2.0$ & $2\,10^5$  & $101$ \\
WA &$6$  & $5$ &20 & $0.2$     & $2.0$ & $4\,10^5$  & $101$ \\
WA &$8$  & $10$ &30 & $0.2$     & $2.0$ & $1.2\,10^6$  & $101$ \\
WA &$10$ & $10$ &40 & $0.2$     & $2.0$ & $6\,10^6$  & $101$ \\
WA &$12$ & $20$ &40 & $1/3$     & $2.0$ & $5\,10^6$  & $101$ \\
\hline
\hline
\end{tabular*}
\end{table}

\subsubsection{Detailed comparison of a single hard sample}

In this section, I make a detailed comparison for the hardest instance of $L=8$ chosen from a total of 2001 instances in Ref.~\cite{TC}. I will focus on the systematic error of $-\beta F$, the $q$ distribution at a low temperature $\beta=5$ as well as the evolution of $\{p_i\}$ as a function of $\beta$. One can see from the evolution of $\{p_i\}$ that this sample is rather chaotic with multiply boundary condition crossings.

\begin{figure}[htb]
\begin{center}
\includegraphics[scale=0.6]{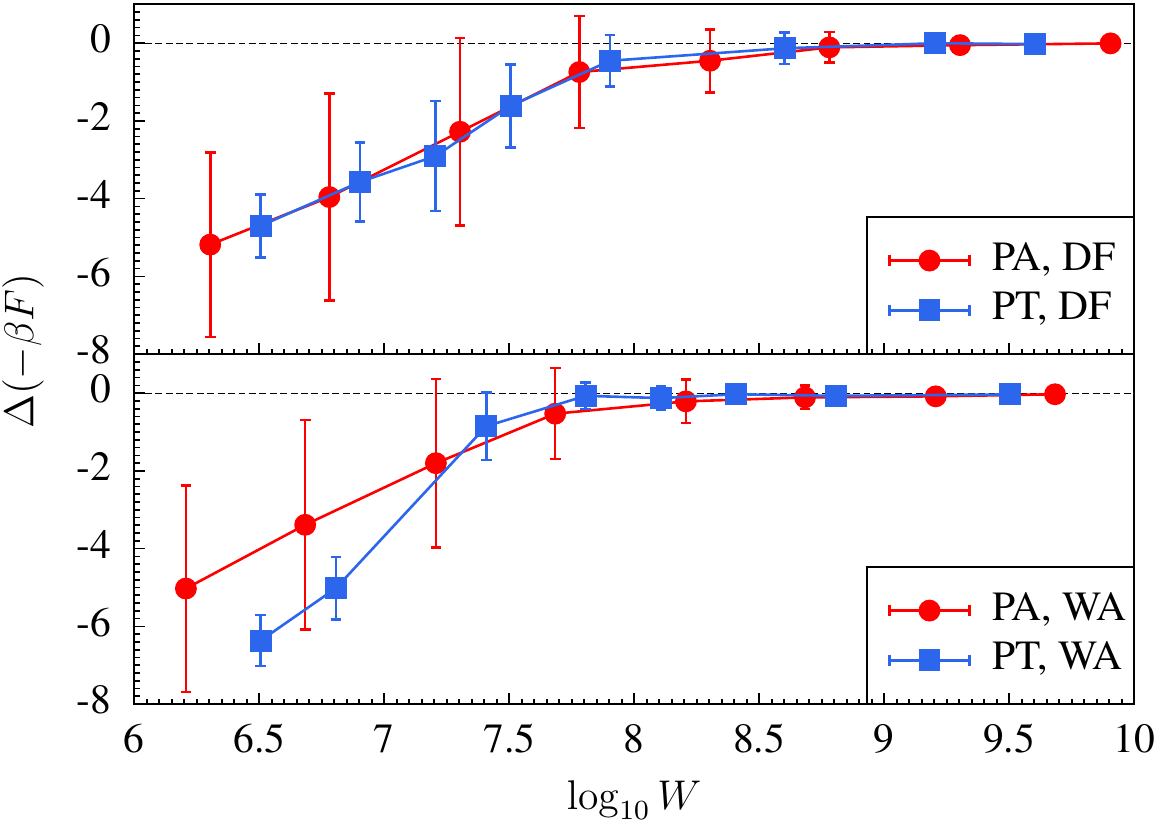}
\caption{(Color online)
Linear-log plot of the systematic error $\Delta(-\beta F)$ of PA and PT as a function of the amount of work $W$ for a chaotic instance of $L=8$ at $T=0.2$. Top panel: the diffusion method (DF). Bottom panel: the weighted average method (WA). The errorbars are the standard deviation of the $-\beta F$ distribution computed from multiple independent runs, not the errorbar of the sample mean of $-\beta F$. Note that even though the weighted average simulates all the 8 boundary conditions independently, it still converges faster than the diffusion method and is therefore more efficient.}
\label{papt}
\end{center}
\end{figure}

\begin{figure}[htb]
\begin{center}
\includegraphics[scale=0.6]{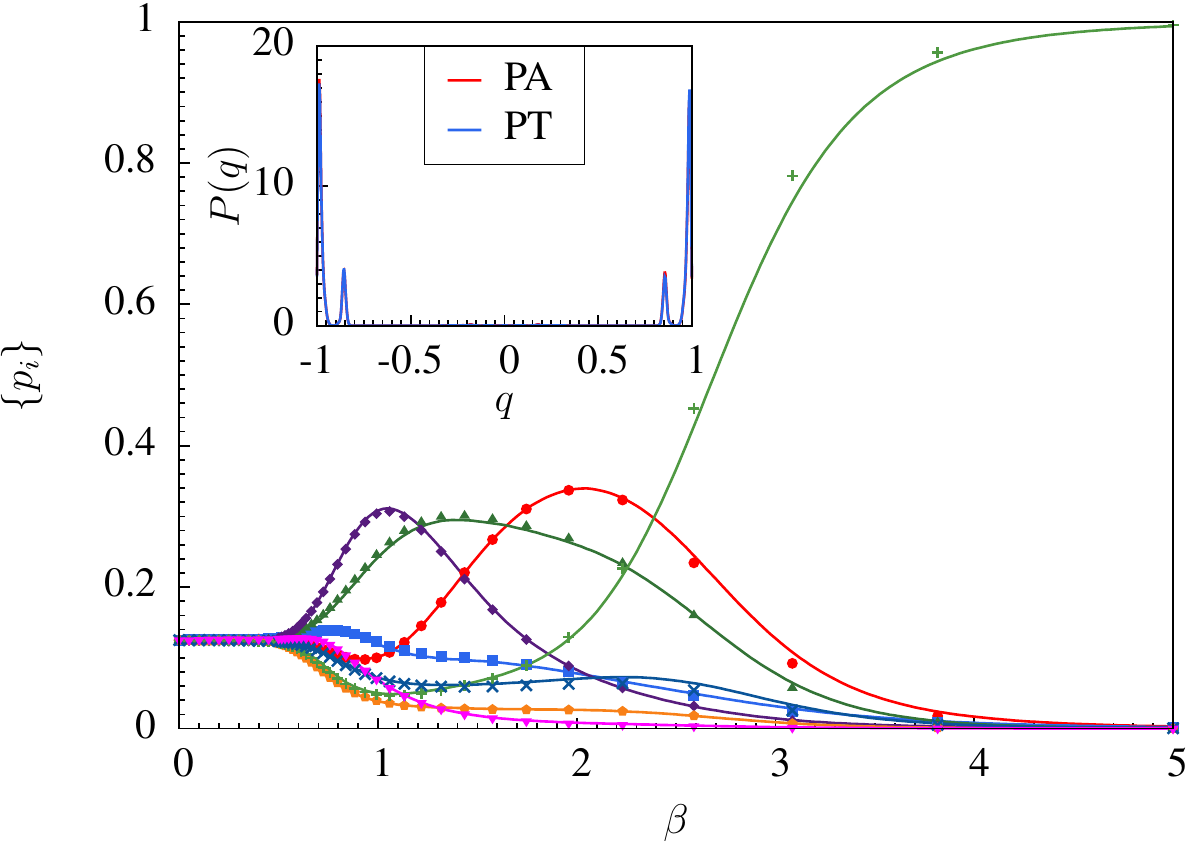}
\caption{(Color online)
The comparison of the evolution of $\{p_i\}$ as a function of $\beta$ of PA and PT using the diffusion method. The solid lines are for population annealing and the points are for parallel tempering. Each color represents a boundary condition. The inset is the comparison of the overlap distribution at $\beta=5$. The results are randomly chosen from the largest runs of Fig.~\ref{papt}.
}
\label{PI}
\end{center}
\end{figure}

The convergence of the dimensionless quantity $-\beta F$ at a low temperature $\beta=5$ is studied as a function of the amount of work $W$. Even though two independent sets of replicas are simulated in parallel tempering to compute the overlap distribution and data were collected from both sets, when studying the convergence of the free energy, this factor of 2 is not counted since this only improves statistical errors, not systematic errors. Indeed, measuring free energy does not require two sets of independent runs. One can say that it is an advantage of population annealing that can effectively compute the overlap distribution from one run due to the ensemble nature of the algorithm. The same parameters in Tables~\ref{para1} and \ref{para2} are used for this instance and the amount of work is varied by changing the population size in PA and the number of sweeps in PT. The convergence of $-\beta F$ is shown in the top panel of Fig.~\ref{papt}. The standard is taken from a weighted average \cite{machta:10} of the largest data set of PA, and the errorbars in the plot are estimated from multiple independent runs. The overlap distribution and the evolution of $\{p_i\}$ as a function of $\beta$ are shown in Fig.~\ref{PI}. The results are randomly chosen from the largest runs of PA and PT in Fig.~\ref{papt}. We see that PA and PT have similar performance in the diffusion method.

\subsubsection{A large-scale comparison}
\label{lsc}

\begin{figure}[htb]
\begin{center}
\includegraphics[scale=0.8]{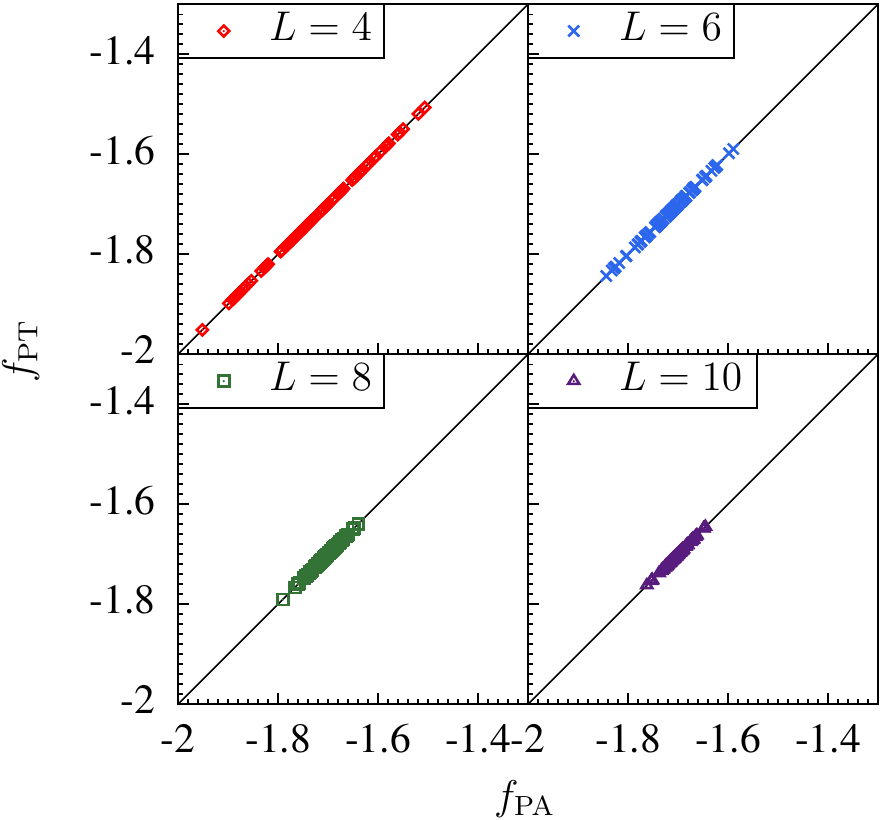}
\caption{(Color online)
Comparison of the free energy per spin $f$ at $\beta=5$ of population annealing and parallel tempering using the diffusion method.}
\label{fs}
\end{center}
\end{figure}

\begin{figure}[htb]
\begin{center}
\includegraphics[scale=0.8]{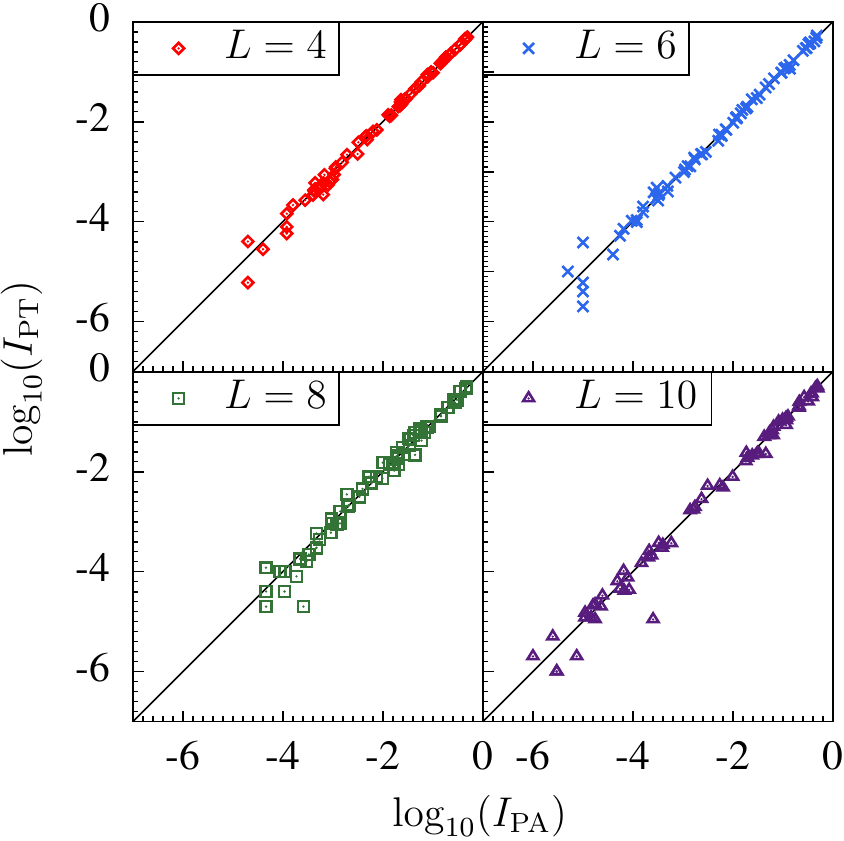}
\caption{(Color online)
Comparison of the statistic $I$ at $\beta=5$ of population annealing and parallel tempering using the diffusion method.}
\label{P0}
\end{center}
\end{figure}

Now we look at a large-scale comparison of PA and PT using the diffusion method. Two important quantities for spin glasses are compared: the free energy per spin $f$ and the statistic $I$ of the overlap distribution. The results for $f$ and $I$ are shown in Figs.~\ref{fs} and~\ref{P0} respectively, again showing the similar performance of PA and PT.

\subsection{Limitations of the diffusion method}
\label{limitation}
We have shown in the last section that one can use the diffusion method to simulate thermal boundary conditions. This method is simple to implement and yields fair result. However, the method has its own limitations due to the diffusion nature of the method, in particular, the problem of the boundary condition dying out.

\begin{figure}[htb]
\begin{center}
\includegraphics[scale=0.6]{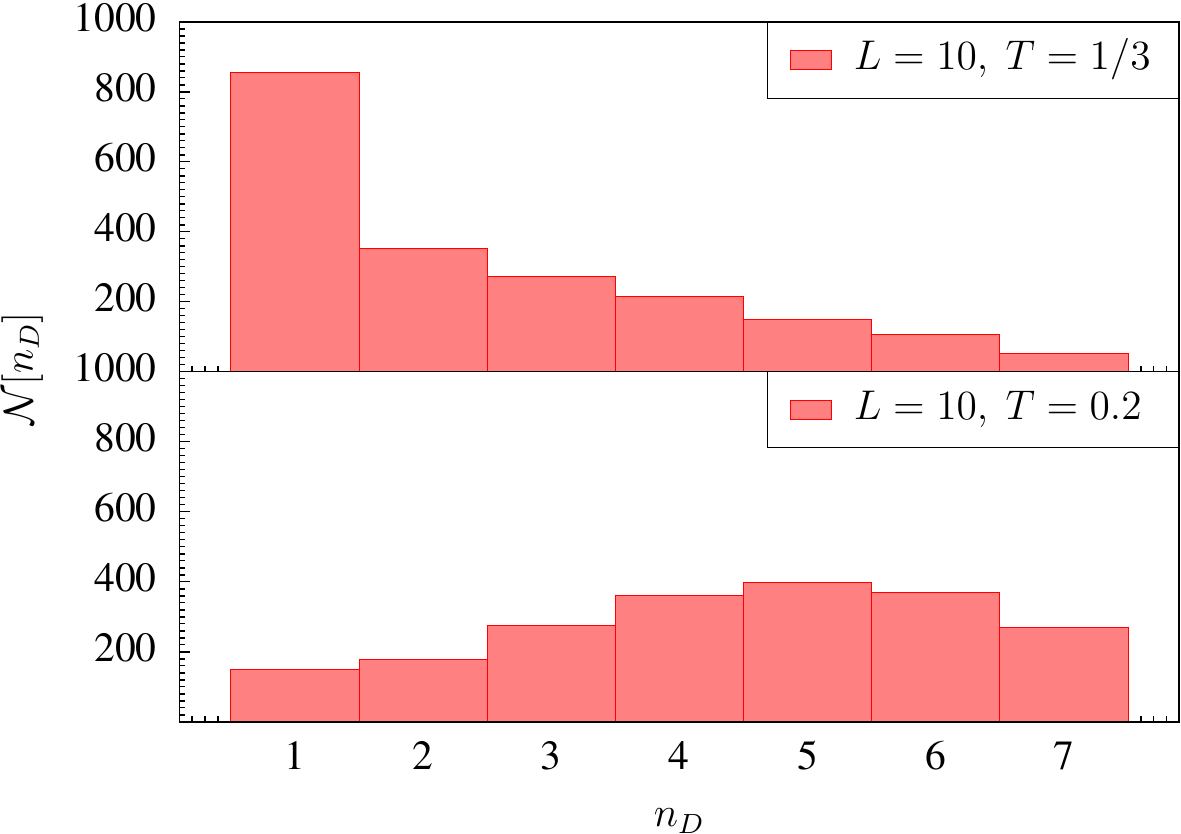}
\caption{(Color online)
Two typical distribution of the number of boundary conditions that die out $n_D$ of $L=10$ at $T=1/3$ and 0.2 respectively in the diffusion method. Note that as temperature is lowered a little in the low-temperature phase, a dramatic change occurs in the distribution as many boundary conditions are eliminated in the annealing process. In contrast, no boundary condition dies out in the weighted average method.
}
\label{Nn}
\end{center}
\end{figure}

The problem of boundary condition dying out of the diffusion method is actually a very severe one. In the state-of-the-art simulation of Ref.~\cite{TC}, where most samples if not all are well equilibrated with decent simulation parameters, it is striking that most of the samples suffer from this problem as shown in Table~\ref{dm} and Fig.~\ref{Nn}. One can see that about $77\%$ of instances have some boundary conditions eliminated at $T=1/3$ and the number quickly grows to about $98\%$ at $T=0.2$. The distribution of the number of boundary conditions that are eliminated $n_D$ also quickly shifts from a bias towards $n_D=0$ to $n_D=7$.

This problem is a resolution problem as the resolution of the weight of a boundary condition is $1/R$. A boundary condition $i$ is likely to be eliminated when $p_i$ becomes close to $1/R$: $p_i \sim 1/R$. When $p_i < 1/R$, it is not even expected to be part of the ensemble in most independent runs. Unluckily, there is no mechanism in the diffusion method to reintroduce a boundary condition once it is eliminated. To ``confirm'' that the simulation is fine, one can only try a much larger $R$, like a factor of 10 as used in Ref.\cite{TC}, and see if the results remain the same. But this is not satisfactory as the runs can be consistently wrong and this is also not efficient. This is especially the case for large system sizes as temperature chaos asserts that the dominate boundary conditions switches more frequently as system size increases. The problem of the total absence of some boundary conditions also occurs in the parallel tempering implementation. Bottlenecks in the diffusion channel prevents some boundary conditions to reach the lowest temperature. This is not too surprising considering the similar performance of PA and PT. Therefore, this is a limitation of the diffusion method itself, independent of which algorithm one uses to implement it. In the next section, I study the weighted average method, its validity as well as its efficiency.

\begin{table}
\caption{
The fraction of instances with boundary condition dying out $f_\mathcal{J}$ in the large-scale simulation of Ref.~\cite{TC} with $M=2001$ samples of each system size $L$. In contrast, no boundary condition dies out in the weighted average method.
\label{dm}
}
\begin{tabular*}{\columnwidth}{@{\extracolsep{\fill}} l c c c c c c l l}
\hline
\hline
$L$  & 4 & 6 & 8 & 10 & 10 & 12 \\
\hline
$T$  & 0.2 & 0.2 & 0.2 & 0.2 & $1/3$  & $1/3$ \\
$f_\mathcal{J}$  & 0.9835 & 0.9840 & 0.9820 & 0.9720 & 0.7701  & 0.7731 \\
\hline
\hline
\end{tabular*}
\end{table}

\subsection{PA and PT: the weighted average method}
\label{m2}

In this section, I present a detailed study of the same hard instance as well as a large-scale simulation using the weighted average method with PA and PT. At first sight, this might be less efficient as one has to simulate all the 8 boundary conditions independently. However, as we shall see in this section that this is not the case. For the hard instance studied, it is even more efficient than the diffusion method. In addition, the problem of boundary conditions elimination is completely removed in the weighted average method.

\subsubsection{Detailed comparison of the single hard sample}
\begin{figure}[htb]
\begin{center}
\includegraphics[scale=0.6]{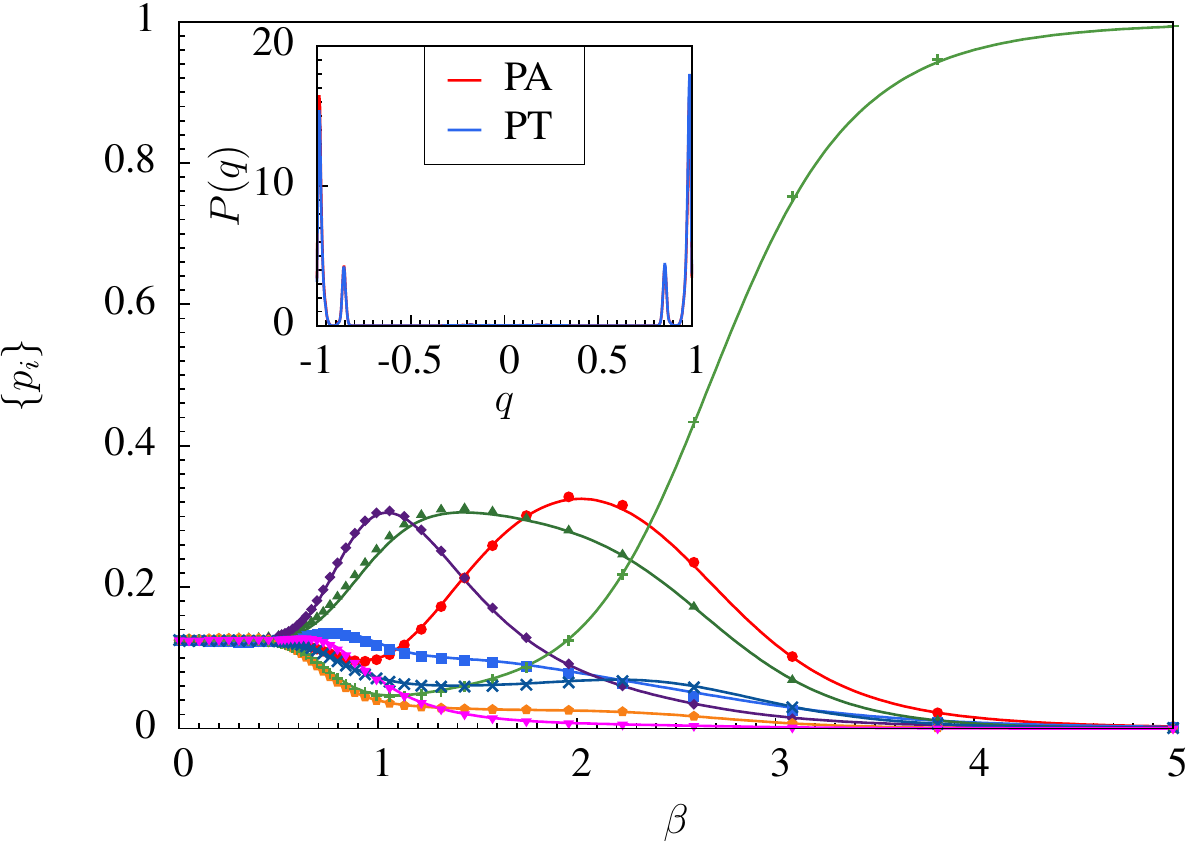}
\caption{(Color online)
The comparison of the evolution of $\{p_i\}$ as a function of $\beta$ of PA and PT using the weighted average method. The solid lines are for population annealing and the points are for parallel tempering. The results are randomly chosen from the largest runs of Fig.~\ref{papt}. The inset is the comparison of the overlap distribution at $\beta=5$. Note that even with slightly less computational work, the agreement is better than that of Fig.~\ref{PI} using the diffusion method, showing that the weighted average method is more accurate.
}
\label{PI2}
\end{center}
\end{figure}

The convergence of $-\beta F$ is shown in the bottom panel of Fig.~\ref{papt}. Compare with the diffusion method in the top panel, one can see that it is more efficient. This is interesting and therefore the weighted average method is both more efficient and more accurate. PA and PT are still comparable in performance. One may have noticed that PT converges slightly faster, but there are more factors to consider here. One is that PT often requires two independent sets of replicas, which are not counted here, to compute the spin overlaps. Take this into account, PA and PT are similar in performance. Another factor is that PT also has slightly more overhead in parallel computing due to the frequent breaks to measure the overlap functions. However, one can choose to run the code in sequential like in the diffusion method. Therefore, it is more fair to conclude that PA and PT are comparable in efficiency. The overlap distribution and the evolution of $\{p_i\}$ as a function of $\beta$ are shown in Fig.~\ref{PI2}. The results are randomly chosen from the largest runs of PA and PT in Fig.~\ref{papt}. The two algorithms give very similar result.

\subsubsection{A large-scale simulation}

\begin{figure}[htb]
\begin{center}
\includegraphics[scale=0.8]{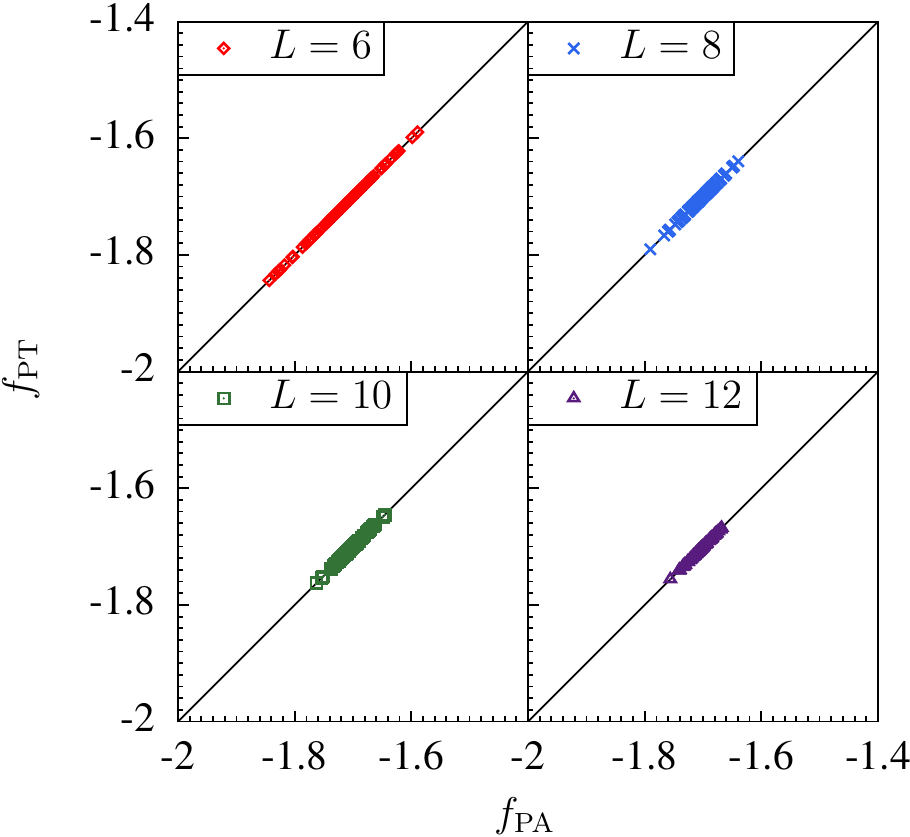}
\caption{(Color online)
Comparison of the free energy per spin $f$ of population annealing and parallel tempering using the weighted average method. $\beta=5$ for $L=6, 8$ and 10 and $\beta=3$ for $L=12$.}
\label{fs1}
\end{center}
\end{figure}

\begin{figure}[htb]
\begin{center}
\includegraphics[scale=0.8]{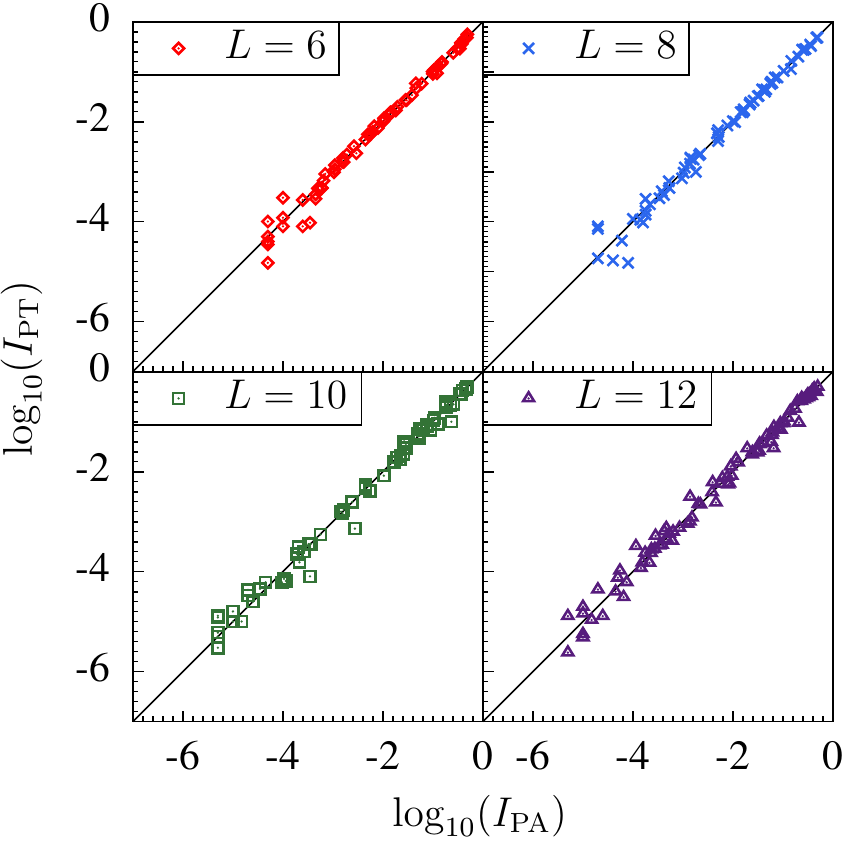}
\caption{(Color online)
Comparison of the statistic $I$ of population annealing and parallel tempering using the weighted average method. $\beta=5$ for $L=6, 8$ and 10 and $\beta=3$ for $L=12$.}
\label{P01}
\end{center}
\end{figure}

Finally, we do a large-scale comparison of PA and PT. We have studied the EA model up to $L=10$ at $T=0.2$ and $L=12$ at $T=1/3$ (current state-of-the-art) using the weighted average method, demonstrating that the method can be used for an accurate large-scale study of spin glasses with thermal boundary conditions. The comparison of the free energy per spin $f$ and the statistic $I$ of the overlap distribution are shown in Figs.~\ref{fs1} and ~\ref{P01} respectively, again showing the similar performance of PA and PT. It also worth noting that the agreement in the measurements of $I$ is better than that the diffusion method even though with slightly less amount of work, in line with the results of the single hard instance.

\section{CONCLUSIONS AND FUTURE CHALLENGES}
\label{conclusion}
In this paper, I studied two methods to simulate thermal boundary conditions, the diffusion method and the weighted average method using both population annealing and parallel tempering. I have illustrated the weaknesses of the diffusion method, in particular, the elimination of boundary conditions. The weighted average method is studied in detail, showing that it is more efficient than the diffusion method and solves the boundary condition elimination problem. Population annealing and parallel tempering on the other hand are similar in their performance.

Furthermore, the weighted method method generates a lot more information than the diffusion method and has opened new ways in doing spin-glass research in thermal boundary conditions. Detailed information of each boundary condition is accessible, and the spin and link overlap distributions between any two boundary conditions are also readily available. For example, it is interesting to study the overlap distributions in the whole instance as well as in a window between the same boundary conditions and two different boundary conditions. This can be very useful in studying the nature of domain walls. i.e. whether the low-lying excitations are space filling or have fractal surfaces. It is also interesting to study the structure of the spin overlap matrix. Except studying the size dependence of the traditional $I$ in periodic boundary conditions and thermal boundary conditions, one can also study the weighted or minimum $I$ of the diagonal part of the overlap matrix. In this way, one can remove effects of the overlaps between different boundary conditions, which are generally larger than the diagonal ones near the origin, and therefore may further reduce finite-size effects. Efforts along these directions are currently in progress and will be presented in future publications.

\section*{Acknowledgments}
I gratefully acknowledge support from NSF (Grant No.~DMR-1208046 and DMR-1151387). I thank Jon Machta for helpful discussions and suggestions, and also for the careful reading of the manuscript. I thank the anonymous referee for insightful suggestions for this manuscript. Finally, I thank the Texas A\&M
University for access to their Ada and Curie clusters.

% use  \cite{Aa10} for references;
%\bibliographystyle{plain}
%\bibliographystyle{apsrevtitle}
%\bibliography{references,refs}

\begin{thebibliography}{15}
\expandafter\ifx\csname natexlab\endcsname\relax\def\natexlab#1{#1}\fi
\expandafter\ifx\csname bibnamefont\endcsname\relax
  \def\bibnamefont#1{#1}\fi
\expandafter\ifx\csname bibfnamefont\endcsname\relax
  \def\bibfnamefont#1{#1}\fi
\expandafter\ifx\csname citenamefont\endcsname\relax
  \def\citenamefont#1{#1}\fi
\expandafter\ifx\csname url\endcsname\relax
  \def\url#1{\texttt{#1}}\fi
\expandafter\ifx\csname urlprefix\endcsname\relax\def\urlprefix{URL }\fi
\providecommand{\bibinfo}[2]{#2}
\providecommand{\eprint}[2][]{\url{#2}}

\bibitem[{\citenamefont{Wang et~al.}(2014)\citenamefont{Wang, Machta, and
  Katzgraber}}]{TBC}
\bibinfo{author}{\bibfnamefont{W.}~\bibnamefont{Wang}},
  \bibinfo{author}{\bibfnamefont{J.}~\bibnamefont{Machta}}, \bibnamefont{and}
  \bibinfo{author}{\bibfnamefont{H.~G.} \bibnamefont{Katzgraber}},
  \bibinfo{journal}{Phys. Rev. B} \textbf{\bibinfo{volume}{90}},
  \bibinfo{pages}{184412} (\bibinfo{year}{2014}).

\bibitem[{\citenamefont{Wang et~al.}(2015{\natexlab{a}})\citenamefont{Wang,
  Machta, and Katzgraber}}]{TC}
\bibinfo{author}{\bibfnamefont{W.}~\bibnamefont{Wang}},
  \bibinfo{author}{\bibfnamefont{J.}~\bibnamefont{Machta}}, \bibnamefont{and}
  \bibinfo{author}{\bibfnamefont{H.~G.} \bibnamefont{Katzgraber}},
  \bibinfo{journal}{Phys. Rev. B} \textbf{\bibinfo{volume}{92}},
  \bibinfo{pages}{094410} (\bibinfo{year}{2015}{\natexlab{a}}).

\bibitem[{\citenamefont{Wang et~al.}(2016)\citenamefont{Wang, Machta, and
  Katzgraber}}]{BC}
\bibinfo{author}{\bibfnamefont{W.}~\bibnamefont{Wang}},
  \bibinfo{author}{\bibfnamefont{J.}~\bibnamefont{Machta}}, \bibnamefont{and}
  \bibinfo{author}{\bibfnamefont{H.~G.} \bibnamefont{Katzgraber}},
  \bibinfo{journal}{ArXiv:1603.00543}  (\bibinfo{year}{2016}).

\bibitem[{\citenamefont{Landry and Coppersmith}(2002)}]{landry:02}
\bibinfo{author}{\bibfnamefont{J.~W.} \bibnamefont{Landry}} \bibnamefont{and}
  \bibinfo{author}{\bibfnamefont{S.~N.} \bibnamefont{Coppersmith}},
  \bibinfo{journal}{Phys. Rev. B} \textbf{\bibinfo{volume}{65}},
  \bibinfo{pages}{134404} (\bibinfo{year}{2002}).

\bibitem[{\citenamefont{Thomas and Middleton}(2007)}]{thomas:07}
\bibinfo{author}{\bibfnamefont{C.~K.} \bibnamefont{Thomas}} \bibnamefont{and}
  \bibinfo{author}{\bibfnamefont{A.~A.} \bibnamefont{Middleton}},
  \bibinfo{journal}{Phys. Rev. B} \textbf{\bibinfo{volume}{76}},
  \bibinfo{pages}{220406(R)} (\bibinfo{year}{2007}).

\bibitem[{\citenamefont{Hukushima}(1999)}]{hukushima:99}
\bibinfo{author}{\bibfnamefont{K.}~\bibnamefont{Hukushima}},
  \bibinfo{journal}{Phys. Rev. E} \textbf{\bibinfo{volume}{60}},
  \bibinfo{pages}{3606} (\bibinfo{year}{1999}).

\bibitem[{\citenamefont{{Sasaki} et~al.}(2005)\citenamefont{{Sasaki},
  {Hukushima}, {Yoshino}, and {Takayama}}}]{sasaki:05}
\bibinfo{author}{\bibfnamefont{M.}~\bibnamefont{{Sasaki}}},
  \bibinfo{author}{\bibfnamefont{K.}~\bibnamefont{{Hukushima}}},
  \bibinfo{author}{\bibfnamefont{H.}~\bibnamefont{{Yoshino}}},
  \bibnamefont{and}
  \bibinfo{author}{\bibfnamefont{H.}~\bibnamefont{{Takayama}}},
  \bibinfo{journal}{Phys. Rev. Lett.} \textbf{\bibinfo{volume}{95}},
  \bibinfo{pages}{267203} (\bibinfo{year}{2005}).

\bibitem[{\citenamefont{{Sasaki} et~al.}(2007)\citenamefont{{Sasaki},
  {Hukushima}, {Yoshino}, and {Takayama}}}]{sasaki:07b}
\bibinfo{author}{\bibfnamefont{M.}~\bibnamefont{{Sasaki}}},
  \bibinfo{author}{\bibfnamefont{K.}~\bibnamefont{{Hukushima}}},
  \bibinfo{author}{\bibfnamefont{H.}~\bibnamefont{{Yoshino}}},
  \bibnamefont{and}
  \bibinfo{author}{\bibfnamefont{H.}~\bibnamefont{{Takayama}}},
  \bibinfo{journal}{Phys. Rev. Lett.} \textbf{\bibinfo{volume}{99}},
  \bibinfo{pages}{137202} (\bibinfo{year}{2007}).

\bibitem[{\citenamefont{Hasenbusch}(1993)}]{hasenbusch:93}
\bibinfo{author}{\bibfnamefont{M.}~\bibnamefont{Hasenbusch}},
  \bibinfo{journal}{Physica A} \textbf{\bibinfo{volume}{197}},
  \bibinfo{pages}{423} (\bibinfo{year}{1993}).

\bibitem[{\citenamefont{Hukushima and Iba}(2003)}]{hukushima:03}
\bibinfo{author}{\bibfnamefont{K.}~\bibnamefont{Hukushima}} \bibnamefont{and}
  \bibinfo{author}{\bibfnamefont{Y.}~\bibnamefont{Iba}}, in
  \emph{\bibinfo{booktitle}{{The Monte Carlo method in the physical sciences:
  celebrating the 50th anniversary of the Metropolis algorithm}}}, edited by
  \bibinfo{editor}{\bibfnamefont{J.~E.} \bibnamefont{Gubernatis}}
  (\bibinfo{publisher}{AIP}, \bibinfo{year}{2003}), vol. \bibinfo{volume}{690},
  p. \bibinfo{pages}{200}.

\bibitem[{\citenamefont{Machta}(2010)}]{machta:10}
\bibinfo{author}{\bibfnamefont{J.}~\bibnamefont{Machta}},
  \bibinfo{journal}{Phys. Rev. E} \textbf{\bibinfo{volume}{82}},
  \bibinfo{pages}{026704} (\bibinfo{year}{2010}).

\bibitem[{\citenamefont{Machta and Ellis}(2011)}]{machta:11}
\bibinfo{author}{\bibfnamefont{J.}~\bibnamefont{Machta}} \bibnamefont{and}
  \bibinfo{author}{\bibfnamefont{R.}~\bibnamefont{Ellis}}, \bibinfo{journal}{J.
  Stat. Phys.} \textbf{\bibinfo{volume}{144}}, \bibinfo{pages}{541}
  (\bibinfo{year}{2011}).

\bibitem[{\citenamefont{Zhou and Chen}(2010)}]{zhou:10}
\bibinfo{author}{\bibfnamefont{E.}~\bibnamefont{Zhou}} \bibnamefont{and}
  \bibinfo{author}{\bibfnamefont{X.}~\bibnamefont{Chen}}, in
  \emph{\bibinfo{booktitle}{{Proceedings of the 2010 Winter Simulation
  Conference (WSC)}}} (\bibinfo{year}{2010}), p. \bibinfo{pages}{1211}.

\bibitem[{\citenamefont{Wang et~al.}(2015{\natexlab{b}})\citenamefont{Wang,
  Machta, and Katzgraber}}]{pamc}
\bibinfo{author}{\bibfnamefont{W.}~\bibnamefont{Wang}},
  \bibinfo{author}{\bibfnamefont{J.}~\bibnamefont{Machta}}, \bibnamefont{and}
  \bibinfo{author}{\bibfnamefont{H.~G.} \bibnamefont{Katzgraber}},
  \bibinfo{journal}{Phys. Rev. E} \textbf{\bibinfo{volume}{92}},
  \bibinfo{pages}{063307} (\bibinfo{year}{2015}{\natexlab{b}}).

\bibitem[{\citenamefont{Wang}(2015)}]{PTFE}
\bibinfo{author}{\bibfnamefont{W.}~\bibnamefont{Wang}}, \bibinfo{journal}{Phys.
  Rev. E} \textbf{\bibinfo{volume}{91}}, \bibinfo{pages}{053303}
  (\bibinfo{year}{2015}).

\end{thebibliography}

\end{document}